\documentclass[twocolumn,english,prx,showpacs]{revtex4-1}

\usepackage[sort&compress]{natbib}
\usepackage{graphicx}

\newcommand{\eqref}[1]{(\ref{#1})}

\newcommand{\tr}{\mathop{\text{tr}}\nolimits}

\newcommand{\ket}[1]{|{#1}\rangle}
\newcommand{\bra}[1]{\langle{#1}|}

\newcommand{\bracket}[2]{\langle#1|#2\rangle}

\newcommand{\Texp}{{\cal T}\mathop{\text{exp}}\nolimits}

\newcommand{\p}{^{\vphantom{\dagger}}}
\newcommand{\h}{^{\dagger}}

\newcommand{\Up}{{\uparrow}}
\newcommand{\Dn}{{\downarrow}}

\newcommand{\aux}{{\mbox{\tiny aux}}}

\begin{document}

\title{Topological Origin of the Fermion Sign Problem}

\author{Mauro Iazzi}
\author{Alexey A. Soluyanov}
\author{Matthias Troyer}

\affiliation{Theoretische Physik, ETH Zurich, 8093 Zurich, Switzerland}

\begin{abstract}
Monte Carlo simulations are a powerful tool for elucidating the properties of complex systems across many disciplines. Not requiring any \textit{a priori} knowledge, they are particularly well suited for exploring new phenomena. However, when applied to fermionic quantum systems, quantum Monte Carlo (QMC) algorithms suffer from the so-called ``negative sign problem", which causes the computational effort to grow exponentially with problem size. Here we demonstrate that the fermion sign problem originates in topological properties of the configurations. In particular, we show that in the widely used auxiliary field approaches the negative sign of a configuration is a geometric phase that is the imaginary time counterpart of the Aharonov-Anandan phase, and reduces to a Berry phase in the adiabatic limit. This provides an intriguing connection between QMC simulations and classification of topological states. Our results shed clarify  the  controversially debated origin of the sign problem in fermionic lattice models.
\end{abstract}

\maketitle

Interesting phenomena can emerge from simple interactions in many-body systems, but analytic solutions are rare making numerical simulations essential for the investigation of their properties. The Monte Carlo method \cite{Metropolis1949} has had great success due to its benign scaling with the system size. By randomly sampling representative configurations of the system, for example using the Metropolis algorithm \cite{Metropolis-JCP-1953}, the properties of interacting many-body systems can be determined in an unbiased way up to statistical sampling errors that are typically small. In many cases the computational effort only increases polynomially, often linearly or with a small power of the system size.

While Monte Carlo algorithms were originally developed for classical systems, Fermi early on suggested that they could be used to simulate quantum systems by using an imaginary time formulation of the Schr\"odinger equation, as reported in Ref.~\cite{Metropolis1949}. This equation describes classical particles performing a random walk in an external potential. In the infinite time limit the distribution of the particles converges to that of the ground state of the quantum system. Identifying finite imaginary times with the inverse temperature $\beta=1/k_BT$ has led to the path-integral formulation of quantum mechanics, where the partition function $Z={\rm Tr}\exp(-\beta H)$ of the quantum system is mapped into a sum $Z=\sum_c w_c$ over the paths $c$ with statistical weights $w_c$. The stochastic sampling of these paths forms the basis of finite temperature quantum Monte Carlo (QMC) algorithms, which have been widely applied to simulate quantum lattice models~\cite{Handscomb1964,Suzuki1976,Blankenbecler1981,Hirsch1986}, the electronic structure of materials \cite{Ceperley1980,Foulkes2001}, ultracold atoms~\cite{Batrouni2002,Astrakharchik2004}, nuclear matter~\cite{Schmidt1999}, and lattice quantum chromodynamics~\cite{Gupta1988}. 

While in classical systems the Boltzmann weights 
are always positive, in QMC the weight of a path configuration can be negative due to particle statistics or gauge fields \cite{Loh1990}. Negative weight configurations cancel contributions of positive ones, resulting in an exponential increase of statistical errors with system size and inverse temperature (see the Supplementary Material for details). A sign problem hence severely limits the application of QMC methods. While the sign problem is representation-dependent, leaving hope for a solution, it is also nondeterministic polynomially (NP) hard \cite{Troyer2005}. This implies that unless P=NP \cite{Cook-ACM-1971}, which is believed to be highly unlikely, there is no generic solution to the sign problem.

Since the sign problem can nevertheless be solved in specific cases \cite{Chandrasekharan-PRL-1999,Wu-PRB-2005,Chandrasekharan-PRD-2010,Berg-SCI-2012,Huffman-PRB-2014}, one may ask if a broader solution of the sign problem may exist for a restricted class of models, such as electrons with (screened) Coulomb interactions or Hubbard models. The origin of the sign problem in these models has remained controversial for decades and several, so far unsuccessful, attempts at a solution have been made \cite{Blankenbecler1981,Sorella1989,Batrouni1990,Muramatsu1992,Matuttis2001,Matuttis2005,Loh2005,Efetov2009,Efetov2010}. We will show that the origin of the fermion sign problem in common fermionic QMC approaches lies in topological properties of the path configurations and a nonzero Aharonov-Anandan phase picked up during evolution~\cite{Aharonov1987}.

\begin{figure}
 \includegraphics[width=0.23\columnwidth,clip=true,trim=200 100 150 150]{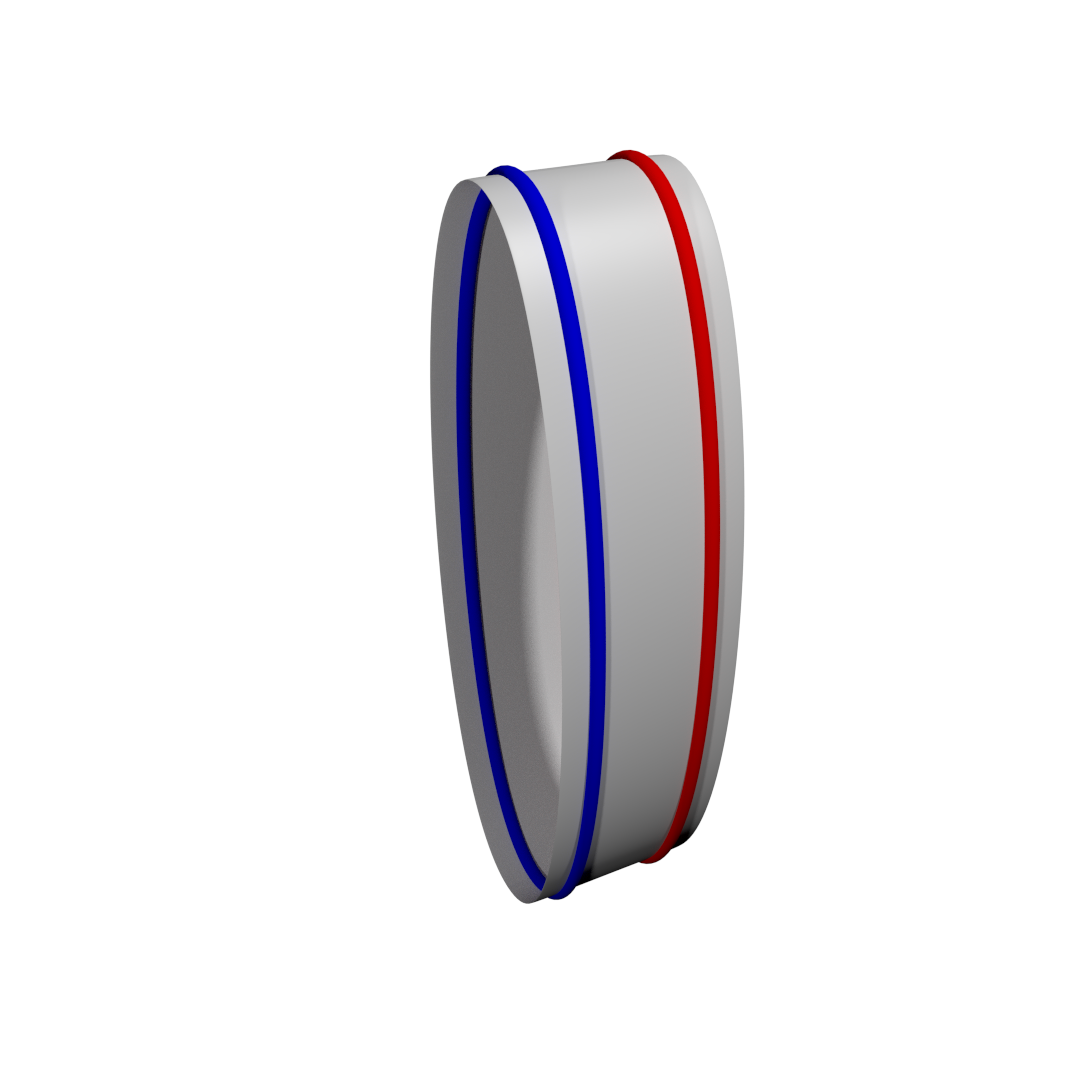}%
 \includegraphics[width=0.5\columnwidth]{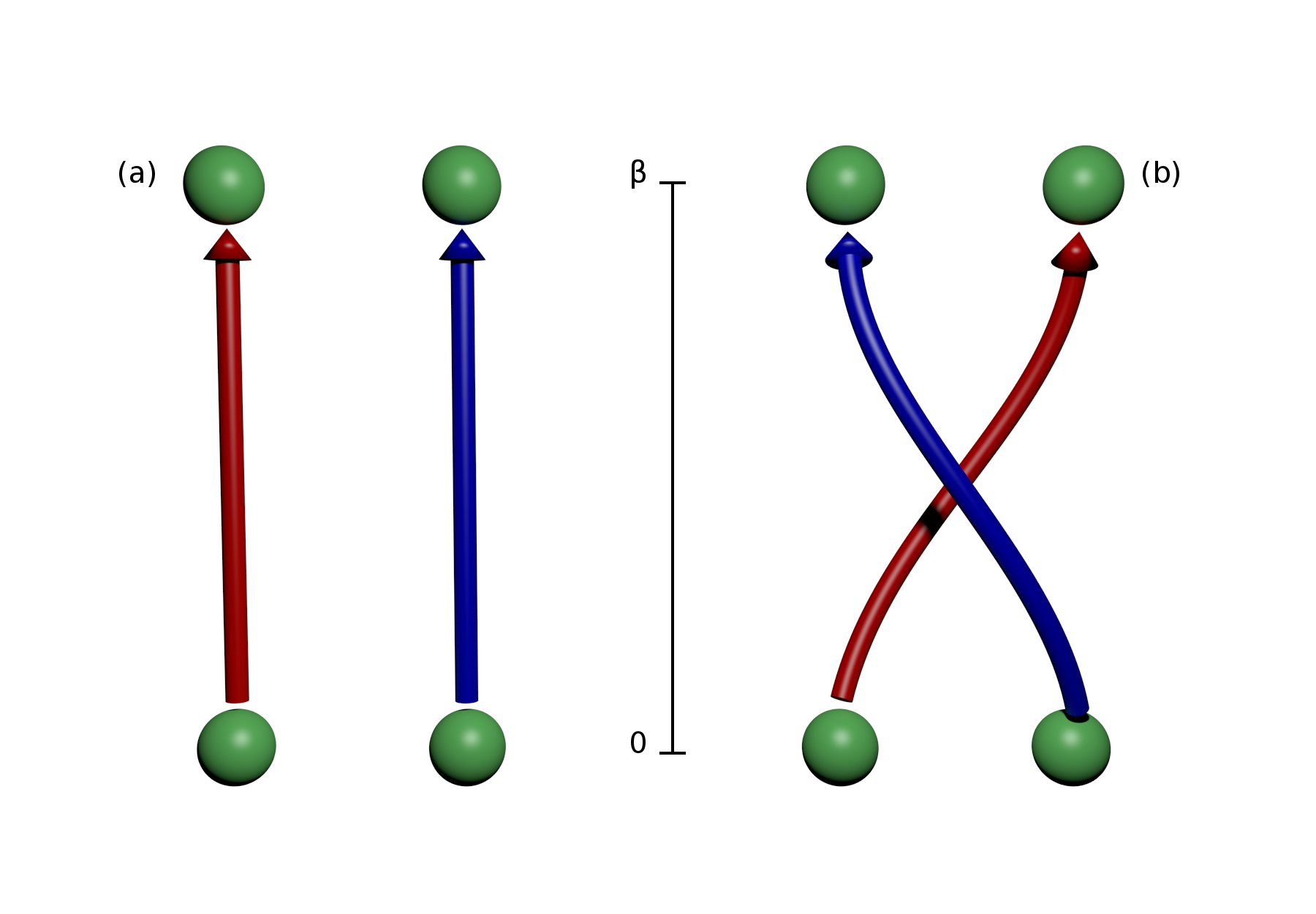}
 \includegraphics[width=0.23\columnwidth,clip=true,trim=200 150 150 150]{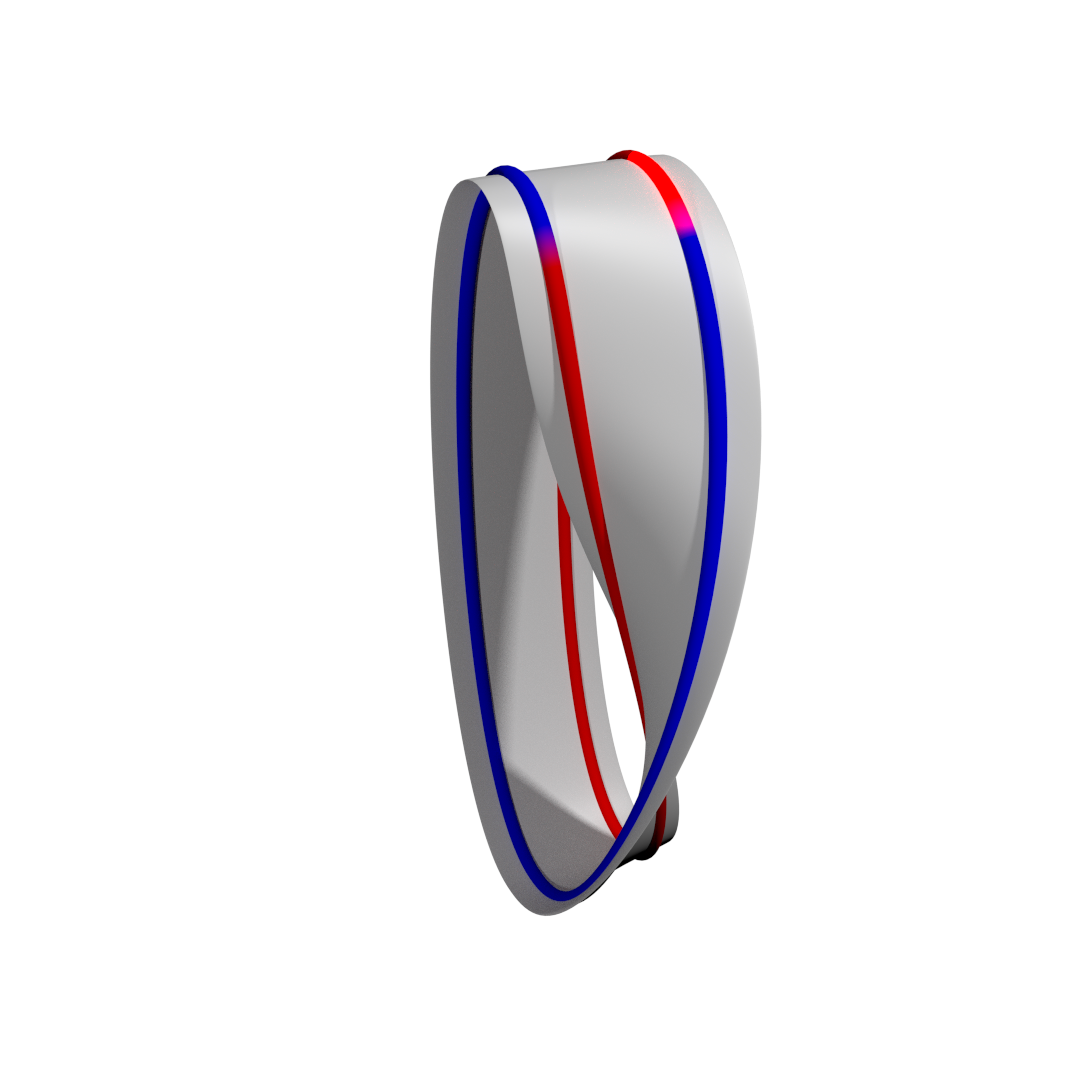}
 \caption{Time evolution in the world-line representation for $2+1$ dimensions. The configuration returns either to itself without exchanging particles and has a positive weight or particles exchange, resulting in a negative sign. The topological nature becomes apparent when connecting the two worldlines by sheet, which closes either to a cylinder or a M\"obius strip.}
\label{fig:exchange}
\end{figure}

The simplest QMC approach is the world line algorithm, which samples the real-space world lines of particles evolving in imaginary time. Indistinguishable particles can be exchanged during the evolution, as is sketched in Fig.~\ref{fig:exchange}. For fermions, a world line configuration with an odd number of exchanges results in a final state that differs in sign from the initial state. After closing the trace it thus contributes with a negative weight. The topological nature of the sign problem, due to the braiding of fermions, is readily apparent. It implies that no local transformation can remove this sign problem but more drastic changes of the representation are needed. One approach to removing this problem is using a \textit{fixed node} method, where an ansatz on the distribution of positive and negative sign regions of phase space is introduced. In select cases such ansatz can be made exact, thus removing any bias introduced by it~\cite{DuBois2014}.

The world line algorithm suffers from a sign problem even for non-interacting fermions. This {trivial} sign problem is solved in an unbiased way by the more efficient and widely used algorithms based on auxiliary fields \cite{Blankenbecler1981,Hirsch1986,Gull2011}. These algorithms map interacting fermions to non-interacting ones coupled to a fluctuating auxiliary field, then integrate out the fermions, to end up with an action of only the bosonic auxiliary field. The origin of the sign problem becomes less transparent in auxiliary field QMC.

To simplify the discussion we focus on the Hubbard model, although our conclusions apply more generally. The Hamiltonian of the Hubbard model 
\begin{equation}
 H = -t\sum_{\langle ij\rangle,\sigma} c\h_{i\sigma}c\p_{j\sigma} -\sum_{i\sigma} \mu_\sigma c\h_{i\sigma}c\p_{i\sigma} + U\sum_{i\sigma} c\h_{i\Up}c\p_{i\Up}c\h_{i\Dn}c\p_{i\Dn}
 \label{eq:hubbard}
\end{equation}
describes fermions with spin $\sigma=\Up,\Dn$, that hop between neighboring lattice sites with a matrix element $t$ and interact via an on-site repulsion $U$, which penalizes double occupancy. The spin-dependent chemical potential $\mu_\sigma$ combines the chemical potential $\mu$ and a Zeeman term. The complete phase diagram of the Hubbard model remains unknown, not the least due to a sign problem of QMC simulations.

To perform auxiliary-field QMC one discretizes the inverse temperature into small imaginary time steps $d\tau$. We use an infinitesimal notation but understand that it refers to both discrete time formulations with finite time steps $\delta\tau$ and the infinitesimal limit \footnote{As always in the definition of path integrals the integrand is well defined only at finite time steps $\delta\tau$. Using the notation of infinitesimals implies that the limit $\delta\tau\rightarrow0$ is taken after performing all evaluations at finite $\delta\tau$}. 

Using a Hubbard-Stratonovich decomposition one rewrites the contribution of each of the interaction terms to $\exp(-d\tau H)$ for one time step as
\begin{equation}
 e^{-d\tau U c\h_{i\Up}c\p_{i\Up}c\h_{i\Dn}c\p_{i\Dn}} \rightarrow \int_{-\infty}^{\infty} d\rho_i e^{ -\frac{\rho_i^2}{2d\tau |U|} + \rho_i(c\h_{i\Up}c\p_{i\Up}+c\h_{i\Dn}c\p_{i\Dn}) },
\end{equation}
where $\rho_i$ represents a component of an auxiliary field, after performing a particle-hole transformation on one spin species in the case of repulsive interactions.
Other choices of auxiliary field decouplings exist but do not to improve the sign problem~\cite{Hirsch1986,Batrouni1990}. A discrete version, which is more \ common in practice~\cite{Hirsch1983}, is discussed in the Supplementary Material. Our conclusions will apply to any such decomposition as well.

After the Hubbard-Stratonovich decomposition we obtain an action that is quadratic in the fermion field operators. This allows one to integrate out the fermion degrees of freedom, obtaining a partition function $Z = \int \mathcal{D}[\bm{\rho}(\tau)] Z_\Up[\bm{\rho}] Z_\Dn[\bm{\rho}]$ that is a path integral over just the auxiliary field configurations. The contribution of a specific configuration $\bm{\rho}(\tau)$ is given by
\begin{equation}
 Z_\sigma[\bm{\rho}] = \det\left[1+e^{\beta\mu_\sigma}\Texp\int_0^\beta d\tau H^{\rm aux}[\bm{\rho}(\tau)]\right],\label{eq:prob}
\end{equation}
where ${\mathcal T}$ indicates time ordering and the chemical potentials $\mu_\sigma$ have been changed by the particle-hole transformation. The matrix $H^{\rm aux}$ is defined through the auxiliary field Hamiltonian
\begin{eqnarray}
 \hat{H}^{\rm aux}[\bm{\rho}(\tau)] &=& -t\sum_{\langle ij\rangle,\sigma} c\h_{i\sigma}c\p_{j\sigma} + \sum_i \rho_i(\tau) \hat{n}_i = \nonumber \\ &=& \sum_{ij,\sigma} H^{\rm aux}_{ij}(\tau)c\h_{i\sigma}c\p_{j\sigma}.
\end{eqnarray}
In certain symmetric cases, for example in the spin-balanced attractive Hubbard model or the half-filled repulsive one, $Z_\Up$ and $Z_\Dn$ have the same sign for each configuration, thus making all weights positive.

The origin of the sign problem in auxiliary field methods has been controversial from the beginning. The original paper \cite{Blankenbecler1981} already suggested that the sign problem should be absent for smooth auxiliary fields. However, unpublished attempts of its authors to remove the sign problem by introducing a smoothing term to the action failed \cite{Scalapino}. On the other hand, it has been suggested that in a ground state projector version of the algorithm a topological  sign problem may exist due to particle exchange similar to the world line algorithm~\cite{Muramatsu1992}. The absence of a sign problem for smooth paths reappeared in recent claims  based on a bosonization approach \cite{Efetov2009,Efetov2010}. Following a different line of argument, it was suggested that the sign problem was merely an artifact of numerical instabilities and could be ignored, while arguing that previous results were unreliable \cite{Matuttis2001,Matuttis2005}. However, these claims were refuted by comparing results on small clusters to numerically exact solutions \cite{Loh2005}.

To investigate the possibility that time discretization or numerical inaccuracies may be the origin of the sign problem we performed simulations using $\delta\tau$ as small as $\frac{t}{200}$, and implementing both a resilient numerical stabilization procedure and a 4096-bit precision version of the algorithm. We found that, while smaller time steps, numerical stabilization and high precision all improved the sign problem there still are configurations where the negative sign remains even after smoothing of the paths. This indicates an intrinsic sign problem even for smooth paths in the continuous time limit.

\begin{figure}[t]
 \begin{center}
  \includegraphics[clip=true,trim=30 40 10 40,width=0.7\columnwidth]{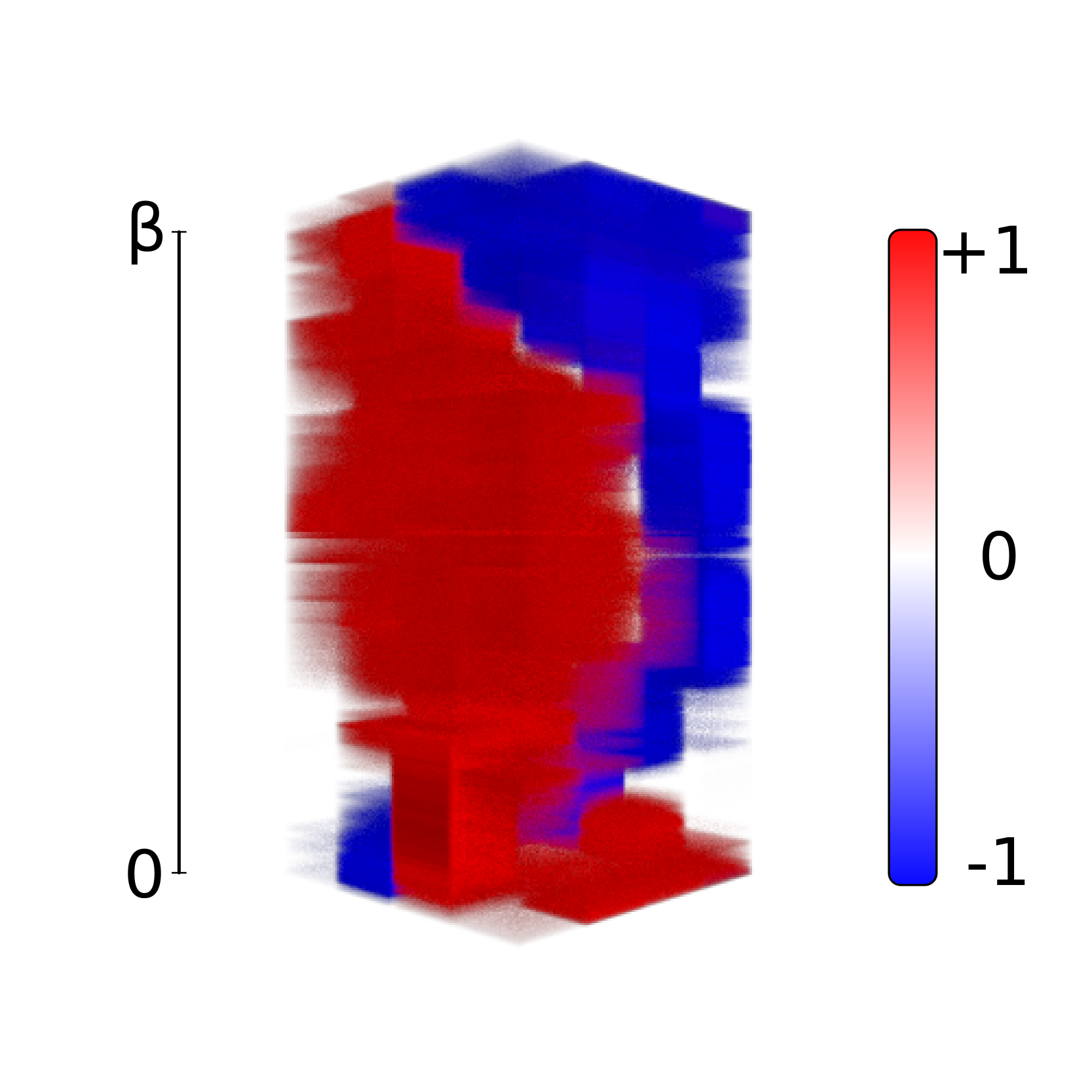}
 \end{center}
 \caption{Imaginary time evolution of the highest occupied state $\ket{\phi_n(\tau)}$ for a smoothed negative-sign configuration on a $4\times4$ Hubbard plaquette. Red color indicates positive-wave function regions and blue indicates negative ones.}
 \label{fig:winding}
\end{figure}

To find the origin of these negative signs we need to discuss the structure of their single fermion modes. While instantaneous eigenvectors $\ket{\psi_n(\tau)}$, defined by 
\begin{equation}
 H_\aux(\tau)\ket{\psi_n(\tau)} = \varepsilon_n(\tau)\ket{\psi_n(\tau)},
\end{equation}
seem intuitive, it is advantageous to instead consider  the \textit{time periodic} eigenstates of the evolution matrix $G(\tau;\beta) = \Texp\int_{\tau}^{\tau+\beta} d\tau' H^{\rm aux}[\bf{\rho}(\tau')]$ and the corresponding eigenvalues
\begin{equation}\label{eq:eig}
 G(\tau;\beta)\ket{\phi_n(\tau)} = \lambda_n\ket{\phi_n(\tau)}
\end{equation}
to understand the sign of a configuration.
We followed the evolution of the eigenvectors of Eq.~\eqref{eq:eig} for negative sign configurations and found that they all contain eigenstates that change sign during time evolution and hence $\lambda_n<0$. The evolution of one of these states is plotted in Fig.~\ref{fig:winding}.  We  see a negative and positive domain winding around each other and finally switching, thus giving a negative weight. 

Expressing the weight of Eq. \eqref{eq:prob}  in terms of the $\lambda_n$ we obtain
\begin{equation}
 Z_\sigma[\bm{\rho}] = \prod_n (1+\lambda_ne^{\beta\mu_\sigma}),
 \label{eq:weight}
\end{equation}
which is negative whenever an odd number of negative-sign single-particle states are more than $50\%$ occupied, {\em i.e.} $|\lambda_n|>e^{-\beta\mu}$. 

\begin{figure}[t]
 \includegraphics[width=\columnwidth]{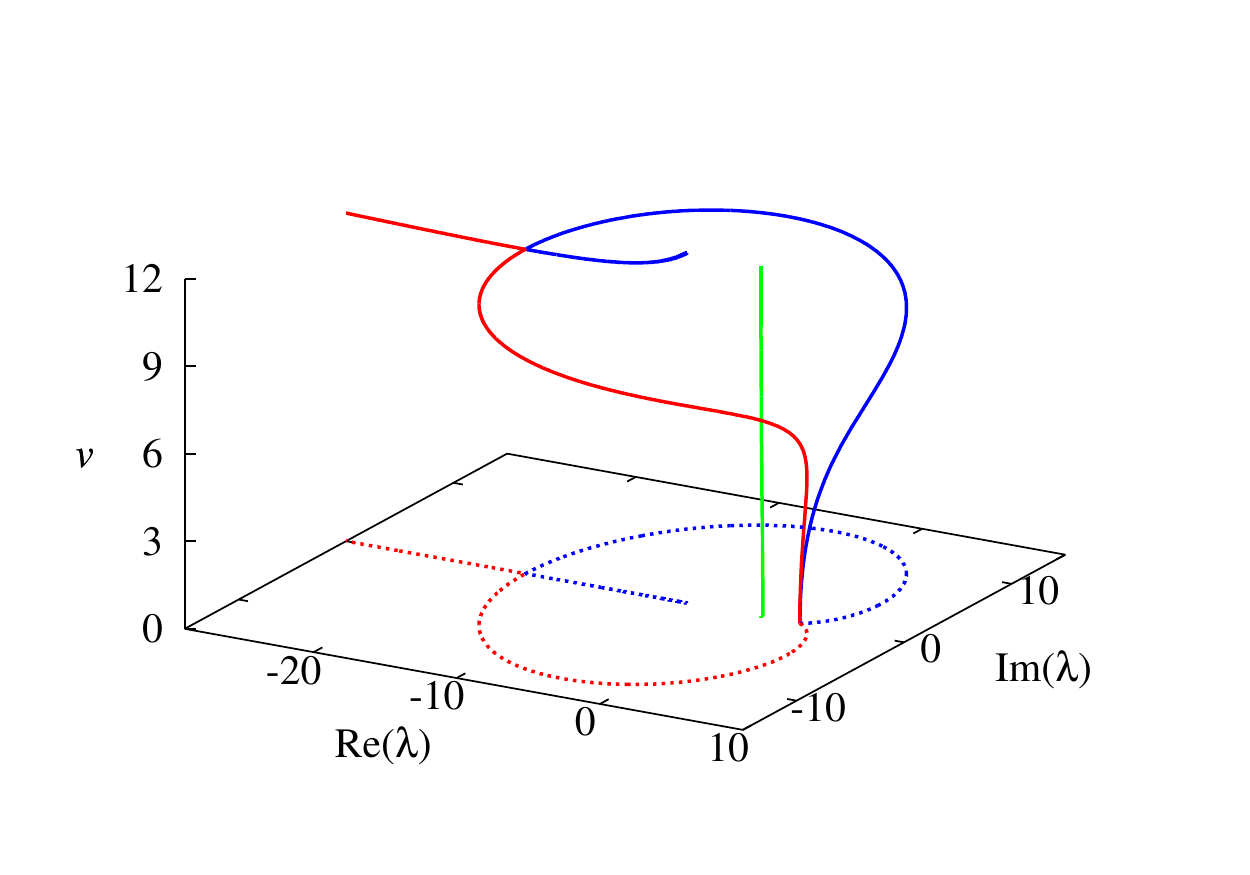}
 \caption{Eigenvalues of the single particle propagator $G(0;\beta)$ with $\beta=1$ as a function of field strength $v$ of the auxiliary Hamiltonian \eqref{eq:three_site}. The dotted lines represent the projections of the paths onto the plane. }
\label{fig:ev}
\end{figure}

After empirically establishing the presence of a sign problem in the Hubbard model we now present a simple and smooth auxiliary field configuration that allows us to understand how the negative signs emerge. The particular configuration we consider is given by the auxiliary field Hamiltonian
\begin{equation}\label{eq:three_site}
 H^{\rm aux}(\tau) = \left(\begin{array}{ccc}
         v\sin(\tau) & -t & -t\\
         -t & v\sin(\tau+2\pi/3) & -t\\
         -t & -t & v\sin(\tau+4\pi/3)
         \end{array}\right),
\end{equation}
which couples a periodic three-site chain to a rotating external field of strength $v$. Similar configurations can be constructed for longer and even length chains. Starting from the non-interacting limit $v=0$, which trivially has positive weights $\lambda_n=e^{-\beta\epsilon_n}$ we now show how negative weights develop when increasing $v$. Since $G(\tau;\beta)$ is a product of real matrices with positive determinant, it is itself real and has positive determinant. Hence no eigenvalue can vanish, complex eigenvalues $\lambda_n$ must come in complex conjugate pairs, and there must always be an even number of negative eigenvalues. These properties were sketched in~\cite{Wu-PRB-2005}. By plotting, in Fig.~\ref{fig:ev}, the eigenvalues of $G(0;\beta)$ as a function of $v$ for $\beta=1$ we see that, indeed, initially all eigenvalues are positive. Ramping up $v$ a doubly degenerate positive real eigenvalue splits into a complex conjugate pair that winds around the $\lambda=0$ line, and rejoins on the negative real axis to a doubly degenerate negative eigenvalue. A sign problem can appear upon further increasing $v$ beyond this critical value $v^*\approx11.2$, when the pair splits into two different real negative eigenvalues, one of the corresponding states becoming occupied and the other unoccupied. Increasing $\beta$ shows $v^*$ approaching zero so that in the zero temperature limit the sign problem is independent of the strength of the auxiliary field and only depends on its geometry.
This winding of pairs of eigenvalues around zero, to become negative, causes the bosonization treatment of Ref.~\cite{Efetov2009} to break down, as we discuss in detail in the Supplementary Material.

The sign of a single-particle state can be understood as a geometric phase by decomposing the eigenvalues in Eq. (\ref{eq:weight}) as $\lambda_n = e^{i\theta_n}\omega_n$ where $\omega_n$ are real positive values. The phases can be computed using the formula
\begin{equation}\label{eq:aaphase}
e^{i\theta_n}= \prod_{\tau=0}^{\beta} \bracket{\phi_n(\tau+d\tau)}{\phi_n(\tau)}.
\end{equation}
These phases are imaginary time versions of the Aharonov-Anandan (AA) phase, which is used to describe the geometric properties of non-adiabatic unitary evolution~\cite{Aharonov1987}.

When the evolution becomes \textit{adiabatic}, which is the case if we stretch a smooth finite temperature configuration by taking $\beta\rightarrow\infty$, the Hamiltonian can be assumed to be locally constant and $e^{-\delta\tau H(\tau)}$ projects onto the instantaneous eigenvectors. Hence in the adiabatic limit $\ket{\psi_n(\tau)}=\ket{\phi_n(\tau)}$. We can then obtain the weights as $w_n=\exp[-\int_0^\beta\varepsilon_n(\tau)d\tau]$ and Eq. \eqref{eq:aaphase} reduces to a Berry phase of the instantaneous eigenstates $\ket{\phi_n(\tau)}$.

The weight $\omega_n$ either diverges or vanishes in the $\beta\rightarrow\infty$ limit, which means that a level is either fully occupied or completely empty, as expected in the case of zero thermal fluctuations.  
The {\em global} geometric phase $\theta$  determining the sign of the configuration is then obtained as the sum of the individual AA phases $\theta_n$ of the occupied levels, which in turn depend only on the geometric properties of the auxiliary field.

As we have seen, $\theta$ can be nonzero even for smooth field configurations, a fact that was missed by Refs. ~\cite{Blankenbecler1981,Efetov2009}. Since the auxiliary Hamiltonian is real, and complex eigenvalues of $G(\beta)$ are always degenerate and thus do not contribute to the overall phase of the configuration, we can limit our treatment to real $\lambda_n$. For these eigenvalues, the wavefunctions $\ket{\phi_n(\tau)}$ can be chosen to be real at all times $\tau$, implying that the AA phases $\theta_n$ are quantized to $0$ or $\pi$. The phases vanish if the wave function can be chosen real and singlevalued, {\it i.e.} $\psi(\tau)=\psi(\tau+\beta)$ at all times $\tau$. However, it may not be possible to make such a choice globally continuous $\tau\in [0,\beta]$, and the wave function may change sign during evolution, meaning $\theta_n=\pi$. 

\begin{figure}
\includegraphics[width=0.9\columnwidth]{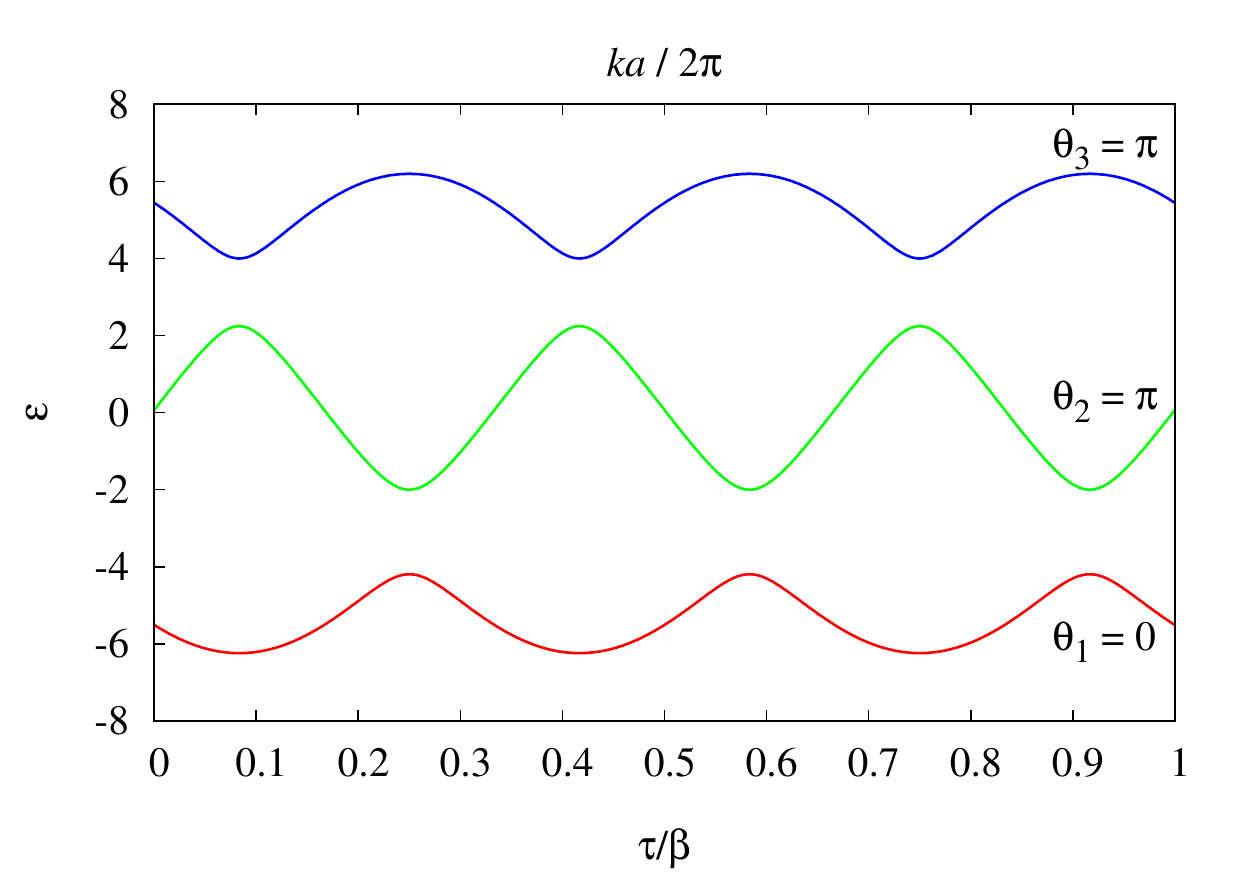}
\caption{Instantaneous levels of the Hamiltonian \eqref{eq:three_site} for $v=6$. They can be interpreted as single particle bands, of which the two top ones have a non-zero Berry phase.}
\label{fig:bloch}
\end{figure}

The quantization of the Berry phase in the adiabatic limit can be understood in terms of a topological invariant associated with the auxiliary Hamiltonian. To make this connection we view the $N$ single particle states of our model as bands of a one-dimensional crystal and relate $\tau$ to the crystal momentum $k$ (and $\beta$ to the primitive reciprocal lattice vector). Each band has a corresponding Berry phase corresponding to the geometric phase accumulated by one-dimensional Bloch states across the Brillouin zone as illustrated in Fig. \ref{fig:bloch}. The relationship of this band structure picture to the AA phase in the case of non-adiabatic evolution is discussed in the Supplementary Material.

Depending on the symmetry of the system, certain topological invariants can be assigned to band structures in question \cite{Schnyder-PRB08}. Such a topological classification of single-particle Hamiltonians is now widely used in the field of topological insulators \cite{Hasan-RMP10}. Protected by symmetry, such topological invariants represent an obstruction for finding globally smooth and momentum-periodic Bloch states that respect the symmetry. In the Hubbard model the real-valued auxiliary field Hamiltonian is responsible for  quantization.

In the presence of an additional symmetry (such as time-reversal) that is preserved by the auxiliary field decomposition the single-particle states come in Kramers pairs. As a consequence, the total winding number vanishes, making the model sign-free \cite{Wu-PRB-2005,Berg-SCI-2012,Huffman-PRB-2014}. The situation here is similar to that of time-reversal symmetric topological insulators \cite{Kane-PRL05}, where the Hilbert space can be split into two subspaces mapped onto each other by time reversal. Each of the sectors can have non zero topological invariants (Chern number) creating an obstruction for choosing a smooth gauge within each of the subspaces. The combined Chern number of the two subspaces, however, is zero \footnote{A smooth gauge is possible in this case but bound not to be time-asymmetric. See Ref.~\onlinecite{Soluyanov-PRB12} for an explicit construction.}. 

It thus turns out that both in the world line picture and the auxiliary field formulation the sign of each configuration is a topological invariant. It hence cannot be simply removed by any local modification or basis change. Understanding the sign of auxiliary field configurations in terms of Aharonov-Anandan phases of the single particle eigenstates, and the connection to topological insulators and superconductors clarifies long-standing open questions about the fermion sign problem \cite{Blankenbecler1981,Muramatsu1992,Samson1993,Efetov2009,Efetov2010}, opens intriguing perspectives for further studies, and provides a path towards the construction of a wider class of sign-problem free models.

We acknowledge discussions with K. Efetov, J. Gukelberger, E. Gull, J. Imri\v{s}ka, A.J. Millis, O. Parcollet, C.~P{\'e}pin, R. Scalettar, and P. Staar. This work was supported by the ERC Advanced Grant SIMCOFE, Microsoft Research, and the Swiss National Science Foundation through the National Competence Centers in Research NCCR QSIT and MARVEL. MT acknowledges hospitality of the Aspen Center for Physics, supported by NSF grant \# 1066293.

\appendix

\section{The negative sign problem}

The foundation of quantum Monte Carlo (QMC) simulation is a mapping of a quantum system to an equivalent classical one by expressing both the partition function
\begin{equation}
Z={\rm Tr}\exp(-\beta H)=\sum_{c\in\Omega} w_c
\end{equation}
and the thermal expectation values of any observable
\begin{equation}
\langle O\rangle = \frac{1}{Z}{\rm Tr}[O\exp(-\beta H)] =  \frac{1}{Z}\sum_{c\in\Omega} O_cw_c.
\end{equation}
as a sum over set of ``classical'' configurations $\Omega$. In the case of a path integral representation this is the set of all path configurations, $c$ is a specific path,  $w_c$ its weight and $O_c$ the contribution of the path to the expectation value of the observable $O$.

For non-negative weights $w_c \ge 0$, this classical system can be sampled by choosing a set of $M$ configurations $\{c_i\}$ from $\Omega$ according to the distribution $w_{c_i}$. The average can then estimated by the sample mean
\begin{equation}
\langle O \rangle \approx \overline{O}=\frac{1}{M}\sum_{i=1}^M O_{c_i},
\end{equation}
within a statistical error 
\begin{equation}
\Delta O = \sqrt{\frac{{\rm Var} O}{M} (2\tau_O+1)},
\end{equation}
where ${\rm Var}O$ is the variance of $O$ and $\tau_O$ is the integrated autocorrelation time of the sequence $\{O_{c_i}\}$.

The standard way of dealing  with the negative weights $w_c$ is to sample with respect to the absolute values of the weights $|w_c|$ and to assign the sign $s_c\equiv {\rm sign}\, w_c$ to the quantity being sampled:
\begin{eqnarray}
\langle O\rangle &=& \frac{\sum_c O_cw_c}{\sum_c w_c} \\ &=& \frac{\sum_c O_cs_c|w_c|\left/\sum_c |w_c|\right.}{\sum_c s_c|w_c|\left/\sum_c |w_c|\right.} \equiv \frac{\langle O s\rangle'}{\langle s \rangle'} .\nonumber
\end{eqnarray}

While this allows Monte Carlo simulations to be performed, the errors increase exponentially with the particle number $N$ and the inverse temperature $\beta$. To see this, consider the mean value of the sign $\langle s \rangle = Z/Z'$, which is just the ratio of the partition functions of the fermionic system $Z=\sum_cw_c$ with weights $w_c$ and the bosonic system used for sampling with $Z'=\sum_c |w_c|$. As the partition functions are exponentials of the corresponding free energies, this ratio is an exponential of the differences $\Delta f$ in the free energy densities \cite{Ceperley1996}: 
\begin{equation}
\langle s \rangle = \frac{Z}{Z'}=\exp(-\beta V \Delta f),
\end{equation}
where $V$ is the volume of the system.
As a consequence, the relative error $\Delta s / \langle s \rangle$ increases exponentially with particle number and inverse temperature:
\begin{equation}
\frac{\Delta s}{\langle s \rangle} 
=\frac{\sqrt{\left(\langle s^2 \rangle-\langle s \rangle^2\right)/M}}{\langle s \rangle} 
=\frac{\sqrt{1-\langle s \rangle^2}}{\sqrt{M}\langle s \rangle} 
\sim 
\frac{e^{\beta V \Delta f}}{\sqrt{M}} .
\end{equation}

Similarly the error for the numerator in Eq.\ (7) increases exponentially and the time needed  to achieve a given relative error scales exponentially in $V$ and $\beta$.

\section{BSS algorithm}

For the discussions in the main paper we focus on the BSS algorithm \cite{Blankenbecler1981}, but note that our results apply more broadly to any auxiliary field algorithm. 
To explain this algorithm in more detail we split the Hubbard Hamiltonian into noninteracting and interacting parts
\begin{eqnarray}
 H_0 &=& -t\sum_{\langle ij\rangle,\sigma} c\h_{i\sigma}c\p_{j\sigma} -\sum_{i\sigma} \mu_\sigma c\h_{i\sigma}c\p_{i\sigma}, \\
 H_I &=& U\sum_{i\sigma} c\h_{i\Up}c\p_{i\Up}c\h_{i\Dn}c\p_{i\Dn}.
\end{eqnarray}

We then decompose the thermal density matrix  using a Trotter decomposition
\begin{equation}
 e^{-\beta H} = \lim_{N\rightarrow\infty} \left(e^{-\frac{\beta}{N}H_0}e^{-\frac{\beta}{N}H_I}\right)^N.
\end{equation}
coupled with either a continuous Hubbard-Stratonovich transformation
\begin{equation}
 e^{-d\tau U c\h_{i\Up}c\p_{i\Up}c\h_{i\Dn}c\p_{i\Dn}} = \int_{-\infty}^{\infty} d\rho_i e^{ \frac{\rho^2}{2d\tau U} + \rho_i(c\h_{i\Up}c\p_{i\Up}-c\h_{i\Dn}c\p_{i\Dn}) },
\end{equation}
where the auxiliary field $\rho_i$ can take any real value or alternatively a discrete one
\begin{eqnarray}
 e^{-d\tau U c\h_{i\Up}c\p_{i\Up}c\h_{i\Dn}c\p_{i\Dn}} &=& 1 + \gamma c\h_{i\Up}c\p_{i\Up}c\h_{i\Dn}c\p_{i\Dn} \nonumber \\&=& \frac12 \sum_{\sigma_i} (1+\sqrt{\gamma}\sigma_i c\h_{i\Up}c\p_{i\Up})(1-\sqrt{\gamma}\sigma_i c\h_{i\Dn}c\p_{i\Dn})  \nonumber \\&=&
  \frac12 \sum_{\rho_i} e^{-d\tau \rho_i (c\h_{i\Up}c\p_{i\Up} - c\h_{i\Dn}c\p_{i\Dn})}
\end{eqnarray}
where $\gamma=1-e^{-d\tau U}$ and the auxiliary field $\rho_i$ can have the two values $-\ln(1\pm\sqrt{e^{d\tau U}-1})/d\tau$. $\rho_i$ diverges with $d\tau\rightarrow0$, showing its fractal nature.

The partition function can be then rewritten as a sum over all configurations of the auxiliary field
\begin{equation}
 Z = \sum_{\{\rho_i(\tau)\}} \tr\left[ \prod_{\tau=1}^{N} e^{-\frac{\beta}{N}H_0} e^{-d\tau \rho_i (c\h_{i\Up}c\p_{i\Up} - c\h_{i\Dn}c\p_{i\Dn})} \right]
\end{equation}
Since the product in the trace is composed of one-particle operators only (exponents are quadratic in the fermionic fields), the result can be obtained through the determinant of matrices in the single-particle picture
\begin{equation}
 Z = \sum_{\{\rho_i(\tau)\}} \det\left\{ 1 + \prod_{\tau=1}^{N} e^{-\frac{\beta}{N}H_0} e^{-d\tau \rho_i (c\h_{i\Up}c\p_{i\Up} - c\h_{i\Dn}c\p_{i\Dn})} \right\},
\end{equation}
where the operators have been replaced by matrices. This can be written compactly using a time-ordered exponential
\begin{equation}
 Z = \int \mathcal{D}[\rho(\tau)] \det\left\{1+\Texp\int_0^\beta d\tau H_{\mbox{\tiny aux}}[\rho(\tau)]\right\},
\end{equation}
having defined 
\begin{equation}
H_{\mbox{\tiny aux}}[\rho(\tau)] = H_{\rm kinetic} + \sum_i \rho_i(\tau) (\hat{n}_{i,\Up} - \hat{n}_{i,\Dn}).
\end{equation}
Moreover, it can be decomposed into the product of determinants for the up and down spin components
\begin{equation}
 Z = \int \mathcal{D}[\rho(\tau)] Z_\Up[\rho] Z_\Dn[\rho].
\end{equation}
When the two contributions are equal, the algorithm is sign-problem free.

In the case of an attractive potential $U<0$, one can obtain a decomposition of the interacting Hamiltonian with a field that couples to the total number of particles rather than the magnetization
\begin{equation}
 e^{-d\tau c\h_{i\Up}c\p_{i\Up}c\h_{i\Dn}c\p_{i\Dn}} = \sum_{\rho_i} e^{-d\tau\rho_i(c\h_{i\Up}c\p_{i\Up}+c\h_{i\Dn}c\p_{i\Dn})}.
\end{equation}
In this case, and in absence of any magnetic field making the up and down populations imbalanced, we have $Z_\Up=Z_\Dn$ and the sign problem vanishes. The same is true when one looks at the half-filled repulsive case, which is related to the above by a particle-hole transformation.

\subsection{Arbitrary precision algorithm}

To examine the sign of a configuration we implemented the BSS algorithm using arbitrary precision floating point numbers. This required the implementation of a QR algorithm, used to compute the determinant in \eqref{eq:prob} and a double QZ step, to find the eigenvalues of the product $G$. Both algorithms were implemented following Ref. \cite{NumericalRecipes}. The power method was used to calculate eigenvectors for Fig. 2 of the main text. 

The QZ step is a decomposition followed by an exchange of Q and R. The step is performed twice to keep all terms real.
\begin{eqnarray}
 A&=&QR \rightarrow A' = Q^{-1}AQ = RQ = Q'R'  \nonumber \\ \rightarrow A'' &=& Q'^{-1}A'Q' = R'Q'
\end{eqnarray}

Since this high precision algorithm is too slow for actual simulations it was  primarily used to  periodically check for correctness of weights. The algorithm gave the same results in all our tests for $4096$ and $2048$ bits of precision, but lower precision calculations using $1048$  or  $512$ sometimes gave different results, in agreement with observed  condition numbers that were as high as $10^{300}$.

\subsection{Stabilization procedure}

At the core of the BSS algorithm is the calculation of the matrix
\begin{equation}
 G(\beta) = \Texp\int_{0}^{\beta} d\tau' H^{\rm aux}[\bf{\rho}(\tau')].
\end{equation}
in Eq. (\ref{eq:prob}). Using a discrete time formulation with $M$ time steps one has to compute a product of matrices
\begin{equation}
G_i = e^{-\delta_\tau H(\tau_i)}
\end{equation}
with $\delta_\tau = \beta/M$. The configuration weight is then computed as 
\begin{equation}
Z=\det\left(1+\prod_{i=1}^M  G_i\right).
\end{equation}

However, multiplying a string of matrices results, in general, in a very ill-conditioned matrix. As the ratio between the largest and lowest eigenvalue diverges information about the lowest eigenvalues and eigenstates is lost when the ratio between smallest and largest eigenvalues  become of the order of roundoff. Calculations of the determinant of $G(\beta)$ then becomes inaccurate.

\begin{figure}
 \includegraphics[width=0.48\textwidth]{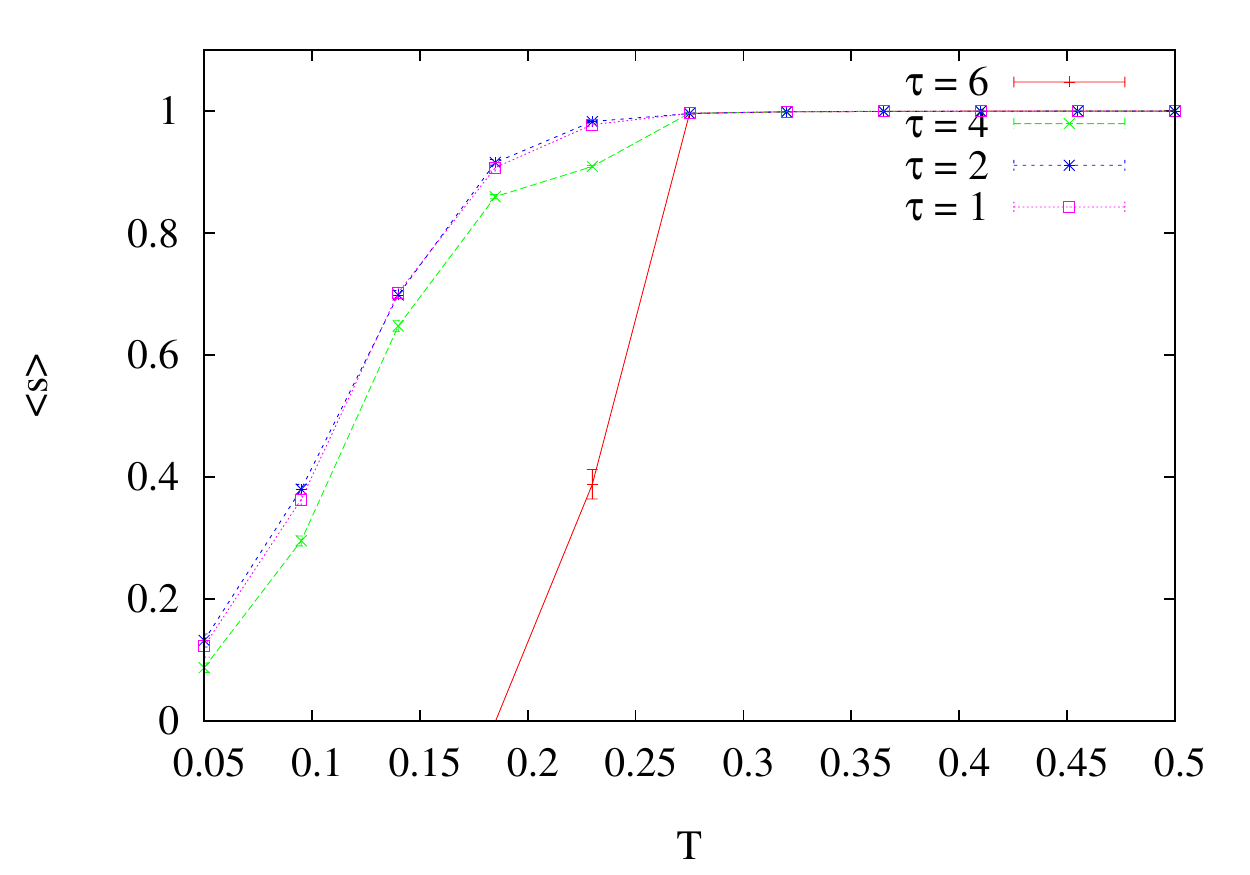}
 \includegraphics[width=0.48\textwidth]{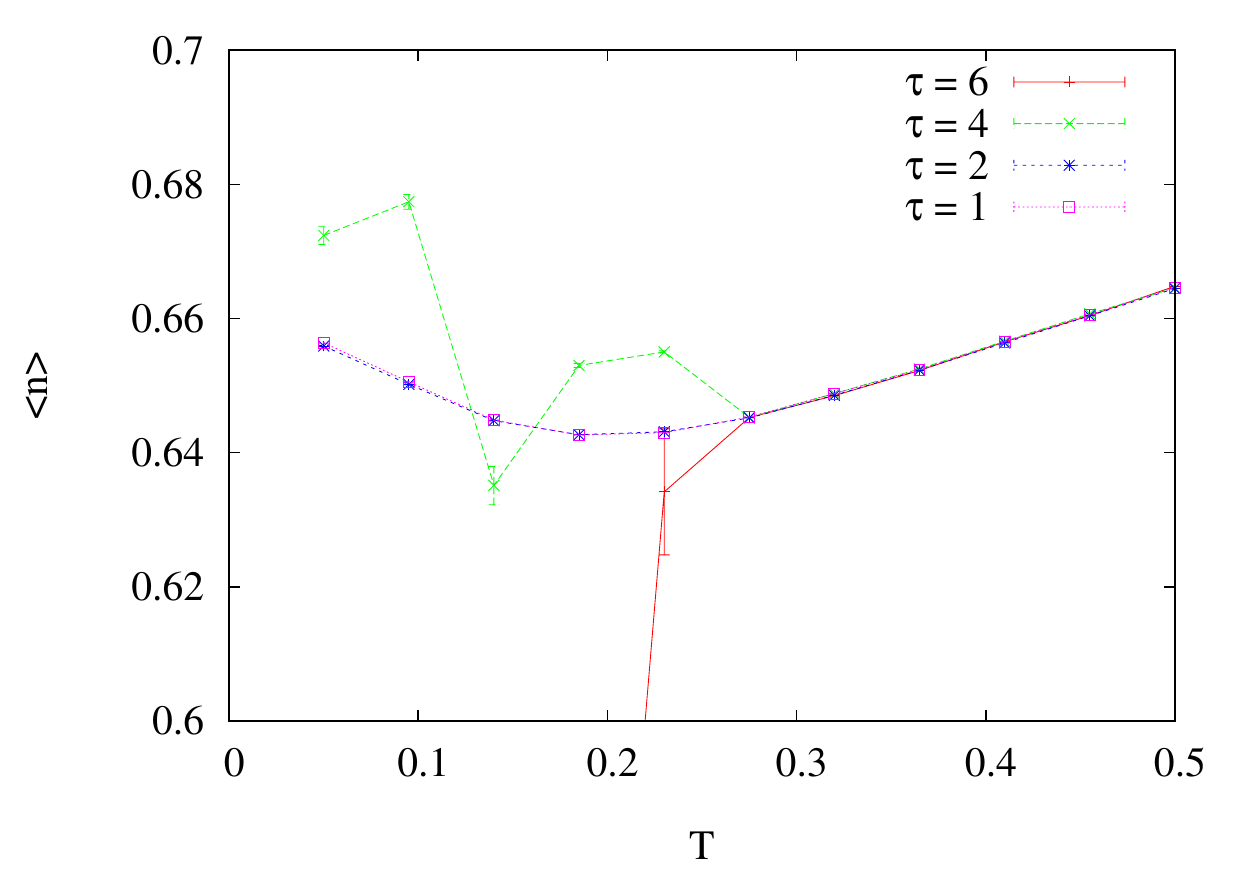}
 \caption{Average sign $s$ and number of particles $n$ at $\mu=0$ ($U/2$ below half filling) as a function of temperature computed with varying distances $\tau$ between SVD decompositions on a $6\times 6$ plaquette. With decreasing temperature, coarser decompositions start to develop a worse sign problem than finer grained ones. This influences even the simplest observables such as particle density.}
 \label{fig:decompositions}
\end{figure}

Numerical stabilization of the product of matrices with an acceptable accuracy is made possible by periodically decomposing the intermediate result using a rank-revealing decomposition such as a singular value decomposition (SVD) or pivoting QR. An SVD is performed on each partial product of a subset consisting of $m$ of the matrices $G_i$, corresponding to an evolution of time $\tau =m\delta_\tau$. We start with
\begin{equation}
\prod_{i=1}^{m} G_i \rightarrow U_k D_1 V_1^T
\end{equation}
The next set of $m$ matrices is then multiplied by $UD$ and decomposed again.
\begin{equation}
\left(\prod_{i=m+1}^{2m} G_i \right) U_1 D_1 \rightarrow U_2 D_2 V_2^T
\end{equation}
The procedure is repeated until the full product has been performed
\begin{equation}
\prod_{i=1}^{M} G_i= U_{M/m} D_{M/m} V_{M/m}^t \ldots V_1^T.
\end{equation}
The value of $m$ (or equivalently $\tau$) should be chosen so that the condition number of the partial products can be stored within machine precision.

We used the arbitrary precision algorithm to check the numerical accuracy of negative sign configurations encountered with different numbers of decompositions . Our results confirmed that while negative signs can creep into the simulation due to numerical errors with a low number of decompositions, such errors disappear with a finer grained stabilization scheme. This is corroborated by the results shown in Fig. \ref{fig:decompositions}

\section{Relationship to prior work}

In this section we discuss in more detail the relationship of our results to prior work, in particular the suggestion of a topological sign problem in projector QMC \cite{Muramatsu1992},  Berry phases in spin models \cite{Samson1993} and the suggestion that the sign problem can be removed by bosonization \cite{Efetov2009}.

We were made aware of an unpublished result by J. Hirsch, where a negative-sign auxiliary field configuration was explicitly constructed to create two localized single particle fermionic states, distant from each other, and then braid them~\cite{Scalettar}. In this configuration there is a clear link between the exchange of particles and the sign. This view however cannot be translated to the general case, where the single particle states are delocalized.

\subsection{Projection Monte Carlo}

The first suggestion of a topological origin of the sign problem appears in the context of  projector quantum Monte Carlo (PQMC), attempting to explain the sign problem similar to the world line algorithm in terms of particle exchange  \cite{Muramatsu1992}. 

 In PQMC one takes the  $T\rightarrow0$ limits, which allows the trace over the thermal density matrix 
\begin{equation}
 \hat{G}(\rho,\beta) = \prod_{\tau=1}^{N} e^{-\frac{\beta}{N}H_0} e^{-d\tau \rho_i (c\h_{i\Up}c\p_{i\Up} - c\h_{i\Dn}c\p_{i\Dn})}
\end{equation}
for a particular auxiliary field configuration $\rho$ to be replaced by a projection from a trial state $\ket{\psi_T}$ (as long as this state is not orthogonal to the ground state):
\begin{equation}
 Z[\rho] = \bra{\psi_T}\hat{G}(\beta)\ket{\psi_T}.
\end{equation}

By choosing a trial state that is a Slater determinant of $p$ fermions described by the $p\times n$ matrix $P$
\begin{equation}
 \ket{\psi_T} = \prod_p (P_{p1}c\h_1 + \ldots + P_{pn}c\h_n)\ket{0}
\end{equation}
we can express the overlap of $\ket{\psi_T}$ with the time evolved state $\hat{G}(\beta)\ket{\psi_T}$ as a determinant in terms of the single particle matrices
\begin{equation}
 \bra{\psi_T}\hat{G}(\beta)\ket{\psi_T} = \det(P^TG(\beta)P)
\end{equation}

If, as is the case for the world-line Monte Carlo, the configuration comes back to itself, {\it i.e.} $\ket{\psi_T}\propto\hat{G}(\beta)\ket{\psi_T}$, it would be clear that each fermion has either come back to its original state, or has exchanged with another particle. In this case the sign would positive or negative depending on the sign of the permutation involved.

This simple picture is complicated in the current case since the time evolved state $\hat{G}(\beta)\ket{\psi_T}$ is , in general, not proportional to $\ket{\psi_T}$. Moreover, it cannot even be written as a Slater determinant of orthogonal vectors as the projection will squeeze the $p$ single particle wave functions that make up $\ket{\psi_T}$ towards the same ground state. 

In Ref.~\cite{Muramatsu1992} a Gram-Schmidt orthogonalization of the vectors  is used to obtain the weight as the determinant of one orthogonal matrix $Q$ times a positive-determinant matrix. The sign then depends on whether $Q$ describes a \textit{proper} or \textit{improper} rotation (i.e. rotation plus reflection). Since the orthogonal matrix can be defined at each time step, after specifying a connection to uniquely identify $Q$ at each step, one can see the evolution as an open curve in the space of orthogonal matrices.

The relationship between this representation of the many-body wave function and the Aharonov-Anandan phases of the single particle states is an interesting topic and deserves additional study.

\subsection{Spin Berry phase}

An interpretation of the sign problem in terms of a spin Berry phase was suggested by Ref.~\cite{Samson1993}. There an auxiliary field decomposition of the Heisenberg Hamiltonian $H_I = J\sum_{\langle ij\rangle}\mathbf{s}_i\cdot\mathbf{s}_j$ is proposed.  Decoupling the spins with an  auxiliary vector field $\mathbf{\Delta}_i(\tau)$ the weight for a given field configuration is expressed as
\begin{eqnarray}
 Z &=& \int \mathcal{D}\mathbf{\Delta} Z[\mathbf{\Delta}]  \\ &=& \int\mathcal{D}\mathbf{\Delta} e^{-\int_0^\beta J^{-1}\sum_{\langle ij\rangle}\mathbf{\Delta}_i(\tau)\cdot\mathbf{\Delta}_j(\tau) d\tau} \nonumber \\ && \times \tr \left[\Texp \int_0^\beta d\tau \sum_i\mathbf{\Delta}_i(\tau)\cdot \mathbf{s}_i(\tau)\right] \nonumber
\end{eqnarray}
The phases gained by the eigenvectors under imaginary time evolution are standard Berry phases of the decoupled spins and may generate a \textit{phase problem} for the spin Hamiltonian.  To our knowledge this paper is the first discussion of a Berry phase in a diffusive (imaginary time) context.

The paper then speculates that a similar Berry phase of spin fluctuations may be the origin of the fermion sign problem in the Hubbard model. However, this relationship is not worked out and in particular because there is no clear relationship between the auxiliary vector field for the decomposition of spin models and the auxiliary scalar field used in fermionic models.  Note also that the auxiliary field approach is not used for spin Hamiltonians, since it generally introduces a phase problem even in models that have no sign problem in a world line formulation. As we have seen, while the origin of the sign problem is also a Berry phase in the Hubbard model, it is not related to the spin Berry phase of Ref.~\cite{Samson1993}.

\subsection{Bosonization}

Ref. \cite{Efetov2009} suggested that bosonization  can be used to remove the sign problem. In their approach the logarithm of the weight of a configuration is written as
\begin{equation}
 \ln Z[\rho] = \int_0^1 \frac{\partial_vZ[v\rho]dv}{Z[v\rho]} + \ln Z[0] + 2\pi in,
\end{equation}
where $n$ is an arbitrary integer, as the phase of $\ln Z$ is only defined up to a multiple of $2\pi$. This can be understood as obtaining the weight $Z[\rho]$ of a configuration starting from the free Hamiltonian $Z[0]$ and slowly ramping up the field strength.  As long as the integral remains real, no sign change can occur in the weight. Whether this is true depends on the behaviour of $1/Z[v\rho]$. Since the integrand is real, it might seem reasonable to assume that the integral is as well -- this however assumes the absence of divergencies.  Following the analysis in the main text, we now explicitly show how such divergencies arise. Making the dependence of the configuration weight on the strength of the auxiliary field $v$ explicit we obtain
\begin{equation}
 Z[v\rho] = \prod_n (1+e^{\beta\mu}\lambda_n(v)),
\end{equation}
for which the bosonization procedure gives
\begin{eqnarray}
  \ln Z[\rho] - \ln Z[0] &=& \sum_n \int_0^1 dv \frac{\lambda_n'(v)}{e^{-\beta\mu}+\lambda_n(v)} \nonumber \\ &=& \sum_n \int_{\gamma_n}d\lambda  \frac{1}{e^{-\beta\mu}+\lambda}
\end{eqnarray}
where $\gamma_n$ is the trajectory of the $n$-th eigenvalue. We can see from our three-site example that one of the eigenvalues crosses the pole at $-e^{-\beta\mu}$ at some value of the field strength $v$. In this case the integrand must be rewritten using the regularization
\begin{equation}
 \frac{1}{x} \rightarrow \mathcal{P} \frac{1}{x} \pm \pi i \delta(x).
\end{equation}
depending on whether it is regularized using the advanced or retarded Green function. Regardless of the choice, the integral will pick up a  contribution $\pi i$ for each eigenvalue crossing the pole, and the weight will be negative when the number of such crossings is odd. As we can see divergencies are common and linked to a change in the Berry phase of the states when increasing the field strength.

A different bosonization scheme was later suggested, which modifies the probability distribution to make all the weights positive~\cite{Efetov2010}. While this method is expected to give different expectation values on finite size lattices, the authors argue that it will converge to the correct values the thermodynamic limit. However, this has not yet been demonstrated.

\section{The adiabatic limit}

Given a configuration $\rho$ at temperature $\beta$, we can relate it to a second one $\rho'$ at $\beta'$ simply by scaling (``stretching'')
\begin{equation}
 \rho'(\tau) = \rho\left(\frac{\beta}{\beta'}\tau\right).
\end{equation}
In the limit $T\rightarrow 0$, the evolution can be taken piecewise constant, so that the element $e^{-\delta\tau H(\tau)}$ essentially projects onto the instantaneous eigenstates.

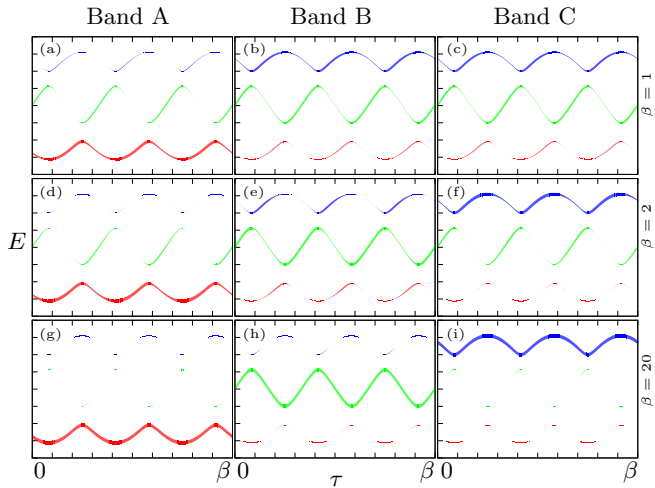
\begin{figure}
\begin{picture}(250,200)
 \put(0,90){$E$}
 \put(9,10){\includegraphics[width=230pt]{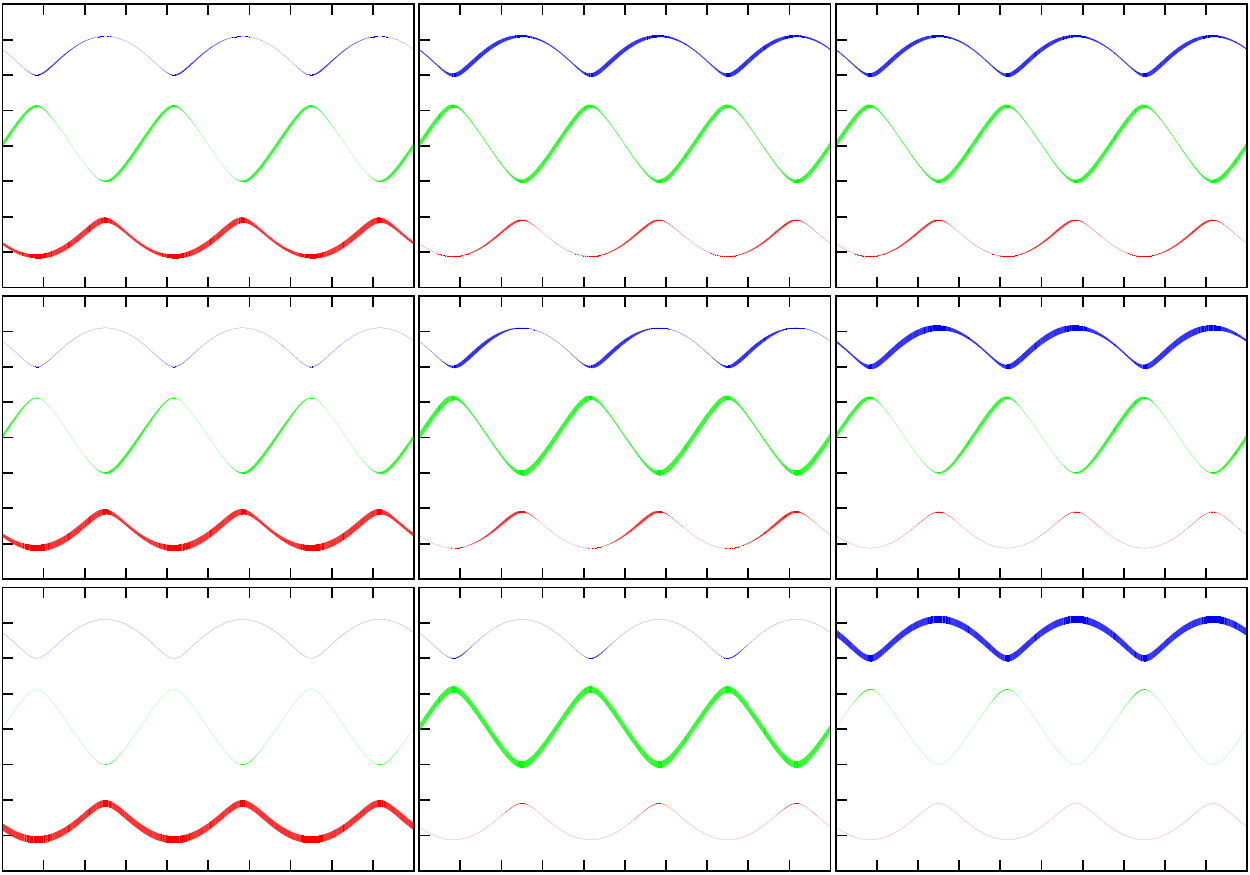}}
 \put(240,140){\rotatebox{90}{\tiny $\beta=1$}}
 \put(240,90){\rotatebox{90}{\tiny $\beta=2$}}
 \put(240,30){\rotatebox{90}{\tiny$\beta=20$}}
 \put(30,175){Band A} \put(107,175){Band B} \put(184,175){Band C}
 \put(12,164){\tiny (a)} \put(89,164){\tiny (b)} \put(166,164){\tiny (c)}
 \put(12,110){\tiny (d)} \put(89,110){\tiny (e)} \put(166,110){\tiny (f)}
 \put(12,56){\tiny (g)} \put(89,56){\tiny (h)} \put(166,56){\tiny (i)}
 \put(10,3){$0$} \put(87,3){$0$} \put(164,3){$0$}
 \put(79,3){$\beta$} \put(156,3){$\beta$} \put(233,3){$\beta$}
 \put(122,0){$\tau$}
\end{picture}
\caption{The periodic eigenstates $\ket{\phi(\tau)}$ of the three band model are shown for different temperature $\beta$ at $v=6$. The line positions  correspond to 
the instantaneous bands of Fig. 4 in the main text, and the line thickness is  proportional to the overlap $|\bracket{\phi_n(\tau)}{\psi_m(\tau)}|^2$. 
At very low temperature (bottom row), the periodic eigenstates follow the instantaneous ones, as the configuration is stretched and the adiabatic approximation becomes valid. The lowest band $A$ is always topologically trivial and does not give rise to a negative sign. At low temperatures the second and third bands $B$ and $C$ carry a $\pi$ Berry phase and are topologically non-trivial. Since only band $B$ is occupied the overall configuration is negative. Raising the temperature (top row) the adiabatic approximation breaks down, bands $B$ and $C$ become degenerate and do not contribute a net phase to the weight. }
\label{fig:bands}
\end{figure}

In Fig. \ref{fig:bands} we illustrate the adiabatic limit for the smooth three-site configuration of the main text. We plot the overlap  overlap $|\bracket{\phi_n(\tau)}{\psi_m(\tau)}|^2$ of the time-periodic eigenstates $\ket{\phi_n(\tau)}$ with the instantaneous ones $\ket{\psi_m(\tau)}$ at three temperatures: above the critical value for the appearance of a sign problem in this configuration,  just below the critical value and at very low temperature. On can observe how at high temperature two periodic eigenstates are degenerate and have the same overlaps. As the eigenvalues $\lambda_n$ split at lower temperature, the overlaps also start to differ and the periodic states pick up a nontrivial phase from the instantaneous states. When the adiabatic approximation becomes valid (last row), each periodic state follows the corresponding band.

\bibliography{biblio}

\end{document}